\documentclass[review]{elsarticle}

\usepackage{color}
\usepackage{comment}
\usepackage{lineno,hyperref}
\journal{Energy}

\begin{document}

\begin{frontmatter}

\title{Thermal Transients in District Heating Systems}
\tnotetext[mytitlenote]{Thermal Transients in District Heating Systems: Physics Modeling for better Control}

\author[MishaLANLddress,MishaSkoltechaddress]{Michael Chertkov \corref{correspondingauthor}
}
\ead{chertkov@lanl.gov}
\ead[url]{https://sites.google.com/site/mchertkov/}
\cortext[correspondingauthor]{Corresponding author}
\address[MishaLANLaddress]{Center for Nonlinear Studies \& T-4, Theoretical Division, Los Alamos National Laboratory, Los Alamos, NM 87545, USA}
\address[MishaSkoltechaddress]{Skolkovo Institute of Science and Technology, 143026 Moscow, Russia}

\author[NNNaddress]{Nikolai N. Novitsky}
\ead{pipenet@isem.sei.irk.ru}
\address[NNNaddress]{Melentiev Energy Systems Institute of Siberian Branch of Russian Academy of Sciences, 130, Lermontov Street, Irkutsk, 664033, Russia}

\begin{abstract}
{\bf
Heat fluxes in a district heating pipeline systems need to be controlled on the scale from minutes to an hour to adjust to evolving demand. There are two principal ways to control the heat flux - keep temperature fixed but adjust velocity of the carrier (typically water) or keep the velocity flow steady but then adjust temperature at the heat producing source (heat plant). We study the latter scenario, commonly used for operations in Russia and Nordic countries, and analyze dynamics of the heat front as it propagates through the system. Steady velocity flows in the district heating pipelines are typically turbulent and incompressible. Changes in the heat, on either consumption or production sides, lead to slow transients which last from tens of minutes to hours. We classify relevant physical phenomena in a single pipe, e.g. turbulent spread of the turbulent front. We then explain how to describe dynamics of temperature and heat flux evolution over a network efficiently and illustrate the network solution on a simple example involving one producer and one consumer of heat connected by ``hot" and ``cold" pipes. We conclude the manuscript motivating future research directions.}
\end{abstract}


\end{frontmatter}

%\linenumbers

\centerline{Highlights}
\begin{itemize}
\item Advection-Diffusion-Loss of heat in district heating networks is analyzed.
\item Parameters of the basic model follow from turbulence phenomenology.
\item Superposition of running and spreading fronts forms a typical transient.
\item We suggest an efficient computational scheme to describe the transients.
\end{itemize}

\section{Introduction}

\subsection{Smart Systems}

Conceptual development of "smart grids" has started in Power Systems (PS) \cite{SmartGridUS,WhatIsSmartGrid}, however extensions of the concept to other energy infrastructures,  such as Natural Gas Systems (NGS) \cite{SmartNaturalGas,SmartNaturalGasEurope,14DCVK} and District Heating (and/or cooling) Systems (DHS) \cite{SmartDistrictHeatingEurope,14PLK}, are now considered as well.  In particular, smart DHS are of a special interest to countries with significant seasonal variations, such as Northern European countries, Russia and potentially USA and Canada. Even countries with milder climate, such as Southern European countries, which do not rely on DHS with a significant (city-scale) footprint, start to reconsider these practices. The city of Torino (Italy) is one impressive example of this type where a very modern system was built during the last 20 years from a scratch \cite{15GSV}.  One may expect, based on this and other recent European examples, that adaptation of this emerging technology to other countries, e.g. USA which so far stay immune to these developments, should be expected soon.

Smartness of a DHS should show itself in how the system is monitored (data), described (physics), controlled (through optimization built on data/measurements and aware of physics) and planned (all of the above). Smart extensions, design and/or re-design of DHS should include long term planning (15-30 years), short term planning (months to years) and operational scheduling/planning (hours to days). Variety of planning and operational problems/formulations should account for uncertainties, take advantage of new technologies and devices which are to be installed in the systems, and build a modern operational version of the system capable to sustain disruptions and emergencies. In combination, these smart developments should be beneficial because of savings, efficiency, convenience and energy security they are expected to bring in. Specific to DHS \cite{10LMMD,14LWWSTHM} one aims to (a) reduce losses by lowering temperature in the system and slowing flows, (b) take advantage of active consumers (demand response), (c) add new smarter devices (pumps, storage, co-generation), (d) rely extensively on sensing and measurements of the mass and heat flows through out the system, (e) co-optimize thermal, electric and gas infrastructures.

\subsection{New challenges}

To advance towards completing the tasks outlined (very schematically) above will face many challenges, in particular:
\begin{itemize}
\item [(1)] Mass transport in the case of an incompressible fluid chosen as the heat carrier (typically water) establishes fast, through a transient propagating with the speed-of-sound, however thermal profile through out the system settles much slower. { Main foundations of the steady mass and heat transport in the DHS were established in 80ies and 90ies, see e.g. \cite{85MK,90SS}. More recently the researchers have focused their attention on analysis of thermal transients  \cite{06GBLS,07GBS,08GBS,08NN,09SZPMKT,17SD}, e.g. addressing the question of how the heat fronts propagate through the system. In these studies the authors investigated how the temperature field,  and thus the heat fronts, are advected  by the flow (velocity) along the pipes. Different effects were considered. In particular, \cite{06GBLS} reported significant dependence of the thermal front spread in response to a rapid change in the flow velocity. It was emphasized in \cite{07GBS} that effects of the systems geometry (especially bends and fittings) influences the thermal front speed and spread rather significantly. Effect of the rapid change of the inlet temperature of the source on the thermal front was analyzed in \cite{08GBS}. Diffusive spread of the front due to turbulence was mentioned and discussed as potentially important in  \cite{06GBLS,07GBS,08GBS}, however it was in the end ignored in the simulations as producing a small effect in the relatively small systems (with spatial extent of 1-2km or less). The turbulent spread of the front was also not accounted in  the simulations of \cite{08NN,09SZPMKT,17SD}. In fact this is not surprising as the systems analyzed in \cite{08NN,09SZPMKT,17SD} were too small for the turbulent diffusion effects to show a significant prominence.

    Even thought it is perfectly reasonable to ignore the turbulent diffusion effects in small systems, or systems where the flows are slow/laminar \footnote{ Note in passing that all the aforementioned studies, reported in \cite{06GBLS,07GBS,08GBS,08NN,09SZPMKT,17SD}, were actually conducted in a turbulent regime with sufficiently large Re-number.}, the turbulent effects should certainly be accounted for in larger systems operated in the turbulent regime (of high Re-number). In other words when the system is turbulent (which is the common case in majority of the district heating systems) and sufficiently large the spread of the front becomes significant enough to cause troubles with timely delivery of services (heat) and with providing sufficiently accurate model for optimal dispatch and control of the heat.}

\item [(2)] Models of DHS should be probabilistic, and as such should account for uncertainty and stochasticity in model parameters, modeling of consumers and producers of heat, and also in modeling of coupling to other infrastructures and (even more generally) other degrees of freedom. Some preliminary work on statistical modeling of DHS was reported by one of the authors of this manuscript \cite{98Nov,08NN,16NV}, however many of relevant statistical questions are yet to be posed and discussed. Note that adapting to DHS statistical and uncertainty-handling methods, recently developed for related problems in power systems, see e.g. \cite{13BLSZZ,14BCH}, is of a great interest.

\item [(3)] Measurements-based monitoring of the system state and model validation are important tasks which were posed and discussed in the literature preliminarily \cite{98Nov}. Further work aimed at developing scalable practical algorithms of both on-line and off-line type resulting in high-quality and reliable predictions is needed.

\item [(4)] The last, but not the least, construction of a comprehensive control, optimization and planning paradigm, built on methods and techniques brought in from other science and engineering disciplines to resolve problems and issues mentioned above, constitutes an over-reaching hyper-challenge.
\end{itemize}

In this paper we mainly focus on addressing the primary challenge \#1, i.e. developing dynamic models { for the heat transport over large systems (of a modern city scale) operating at high Re numbers which are to be used in the future in  tests and validations related to challenges \#2-4.}

\subsection{Operational Modeling}

Most important characteristics of the heat flow spatio-temporal evolution within DHS is that it is never steady. This is in contrast with the mass flow,  which is almost always steady or quasi-steady, under exception of very short transients when it changes and then settles with the speed of sound. It is appropriate to note here that design of the basic principles of the heat flux control has a significant effect on how one models the thermal/heat transients. We distinguish two principally different paradigms. Control-by-temperature which is adopted in Nordic countries and Russia where the first large (city scale) DHS systems were designed a century ago. In these systems mass flow is adjusted rarely, and even then mass flow transients are settled in the matter of seconds and maximum minutes. Main controls are implemented via change of temperature at the heat sources,  that is at the major Combined Heat \& Power (CHP) plants and big boilers. The temperature set points are changed through out the day to adjust to the change in the heat demand. Change of the temperature set point generates a heat front propagating with speed of the mass flow which varies from tenth of meters per second to a few meters per second. Propagation of the heat front through a system (where it does not diffuse yet and is seen as a clear front) may have a significant spatial extend -- from kilometers to tens of kilometers, lasting from minutes to hours. This raises the questions we aim to address in the paper --  of controlling delays in heat delivery and quantifying details of the slow transients. As recognized relatively recently modeling delays, and thus thermal transients while accounting not only for ballistic propagation of the heat fronts but also for diffusive spread of the fronts, is significant \cite{06GBLS,07GBS,08GBS,08NN}. Notice that modern systems build in S. European countries have adopted another operational principe -- control-by-mass flows when the amount of heat flux is controlled through adjustment of the flow velocity \cite{15GSV}.  In such systems flow velocity and heat fluxes are settled (speed of sound) fast. In the heat delivery (hot) part of such systems temperatures are kept constant through out the day,  while in the return (cold) part of the system temperature varies depending on the amount of heat consumed by customers. Therefore propagation and diffusive spread of the heat fronts in such mass-flow-controlled system may be of a significance only in the return (cold) part of the system. Note also that it is natural to expect that the two main operational principles will be merged and combined in smart future DHSs.

\subsection{Layout of Material and Our Main Results}

Layout of material in the manuscript and our main results reported in this manuscript are as follows:
\begin{itemize}
\item We briefly review standard description of the steady hydro (mass) flow in the pipes averaged over turbulent fluctuations, in Section \ref{sec:hydro}.

\item Basic equation describing dynamics of temperature in the steady mass flow is discussed in Section \ref{sec:thermal}. We first introduce the basic thermo-advection-losses equation and then, assuming that the mass flow of the incompressible carrier (water) in the DHS is turbulent, we describe in Section \ref{subsec:turb_phen} how standard turbulence theory phenomenology can be used to argue validity of the basic equation. We present parametric estimates for the turbulent diffusion and heat losses coefficients, both averaged over spatial (cross-section of the pipe) and temporal (cross-section of the pipe times inverse flow velocity) as a function of mean flow velocity. This analysis results in a conclusion that the 1+1 (space+time)- dimensional advection-diffusion-losses equation, supplemented by initial and boundary conditions, are appropriate for modeling thermal transient in an individual pipe and system of pipes. Quantitative estimates for the parameters correspondent to practical DHS system are given in Section \ref{subsec:exemp}. We also discuss estimation for (and importance of) the heat flux in DHS in Section \ref{subsec:heat_flux}.

\item In Section \ref{sec:Cauchi} we solve the advection-diffusion-losses equation, introduced and discussed in Section \ref{sec:thermal}, in the setting of a single pipe. Relevant additional/auxiliary material is also placed in \ref{app:BC_temp}.

\item We discuss extension of the modeling described in Section \ref{sec:Cauchi} for a single pipe to a DHS network in Section \ref{sec:network}, first addressing the static case in Section \ref{subsec:network_static} and then dynamics/transient case in Section \ref{subsec:network_dynamic}. The network solution is illustrated in Section \ref{subsec:two_pipes} on example of the network consisted of one producer of heat and one consumer of heat connected by two pipes, hot and cold.

\item Section \ref{sec:conclusions} is reserved for discussions of the results, conclusions and path forward (future tasks).

\end{itemize}

\section{Hydro (mass-) flow equations}
\label{sec:hydro}

Description of the hydro-thermal flow equations is split in two parts.  First, we describe equations governing mass flows over the system pipes.  Then, we move to description of the temperature field spatio-temporal spread over the system affected by the flow. We will not consider here the speed-of-sound transients which are settled in a matter of seconds (or faster). We will also assume that the heat carrier is a single-phase fluid (water) with the pipes fully filled (no gas component present) - thus ignoring extreme regimes when the water can vaporize or be in the water-vapor (two-phase) regime. Overall, inncompressibility of the single-phase flow with the focus on physical processes, which are slower than the speed-of-sound effects, means that we are in the balanced regime, i.e. in response to consumption change there should be an instantaneous (within the approximation) adjustment of the production. (The situation is quite different in compressible flows, e.g. describing transport of natural gas in the pipes, where the so-called line pack effect, resulting in variability of the total amount of gas contained within the pipes, takes place.) Under these assumptions, the set of balanced quasi-static equations, explaining distribution of the mass-flow and pressure over the entire system (network) of pipes, becomes
\begin{eqnarray}
&& \forall a\in{\cal V}:\quad\underbrace{\varphi_a=\sum_{b\sim a} \phi_{a b}}_{\mbox{flow balance}}\label{Flow1}\\
&& \forall \{a,b\}\in{\cal E}:\quad \underbrace{\frac{p_a-p_b}{L_{ab}}=F(\phi_{ab})}_{\mbox{pressure drop to flow relation}},
\label{Flow2}
\end{eqnarray}
where ${\cal G}=({\cal V},{\cal E})$ is the undirected graph of the pipe network, { and $a\in {\cal V}$ and $\{a,b\}\in{\cal E}$ state that node $a$ and edge $\{a,b\}$ are taken from the set of nodes, ${\cal V}$, and set of edges, ${\cal E}$, respectively}. Other characteristics entering Eqs.~(\ref{Flow1},\ref{Flow2}) are as follows. $\varphi=(\varphi_a|a\in{\cal V})$ is the vector of the mass-flow injections/consumptions which is globally balanced (in view of the comments above), $\sum_{a\in{\cal V}} \varphi_a=0$. $\phi=(\phi_{ab}=-\phi_{ba}|\{a,b\}\in{\cal E})$ is the vector of mass flows over pipes, where each component is odd with respect of the (edge/pipe) index permutations. Eq.~(\ref{Flow2}) describes relation between the pressure drop at the two ends of the pipe, $p_a-p_b$, per length of the pipe, $L_{ab}=L_{ba}$, and the steady mass flow along the pipe, $\phi_{ab}$.  $F(\phi)$ is a nonlinear function representing effect of turbulent friction. According to the standard fluid mechanics estimation,
$F(\phi)\sim \phi|\phi|$, however some other modifications of the simple quadratic dependence \cite{15MN} may be more appropriate to model (phenomenologically) effects of pumps, flow regulators and other control devices.

\section{Temperature (thermal) advection-diffusion equations: the case of a single pipe}
\label{sec:thermal}

Consider a single pipe.  Mass flow is related to density, $\rho$, and (averaged over cross-section) velocity, $V$, according to $\phi=\rho V \pi (d/2)^2$, where $d$ is the diameter of the pipe (assumed constant along the pipe). Assuming also that thermodynamic variations (of density and temperature) are small, the constancy of $\phi$ along the pipe translates into constancy of the flow velocity along the pipe.

Coarse-grained, and already averaged over cross-section of a pipe dynamics of a thermal field in a steady (but turbulent) flow is controlled by the following advection-diffusion equation
\begin{equation}
 \partial_t T+V\partial_x T
 -D \partial_x^2 T +\gamma T=0, %-T_{\mbox{amb}}),
 \label{heat}
 \end{equation}
 where $T(t;x)$ is the temperature field, possibly dependent on time, $t$, and position along the pipe, $x$, and counted from an ambient temperature (here ``ambient" should be counted not as an outside temperature but temperature of the pipe, under insulation coat);
 $D$ is the eddy (turbulent) diffusion coefficient; $\gamma$ is the coefficient of the thermal heat loss through the walls of the pipe (assumed kept at the ambient temperature). $D$ and $\gamma$ are related to $V$, as well as to the geometric parameters of the pipe, and in the turbulent regime can be estimated (through straightforward, a-la Richarson-Kolmogorov-Obukhov-Kraichnan (RKOK), phenomenology) as
 \begin{eqnarray}
 && D\sim V d,\label{eddy_diff}\\
 && \gamma \sim \kappa^{3/4} V^{3/4} d^{-5/4} \nu^{-1/2}\label{eddy_loss}
 \end{eqnarray}
where $\kappa$  and $\nu$ are thermal diffusion coefficient and kinematic viscosity of the water, respectively.
We provide brief description of the RKOK picture of flow and heat profile leading to Eqs.~(\ref{Flow1},\ref{Flow2}) and Eqs.~(\ref{heat}) and resulting in
the estimations of eddy diffusivity (\ref{eddy_diff}) and thermo-losses (\ref{eddy_loss}) respectively, in the next Subsection \ref{subsec:turb_phen}.

Realistic estimates for the eddy-diffusivity (\ref{eddy_diff}) and thermo-loss (\ref{eddy_loss}) coefficients in DHS pipes are given in Subsection \ref{subsec:exemp}.

\subsection{Richardson-Kolmogorov-Obukhov-Kraichnan picture of flow and heat profile in a turbulent pipe}
\label{subsec:turb_phen}

\begin{figure}[t]
\centering
\includegraphics[width=\textwidth]{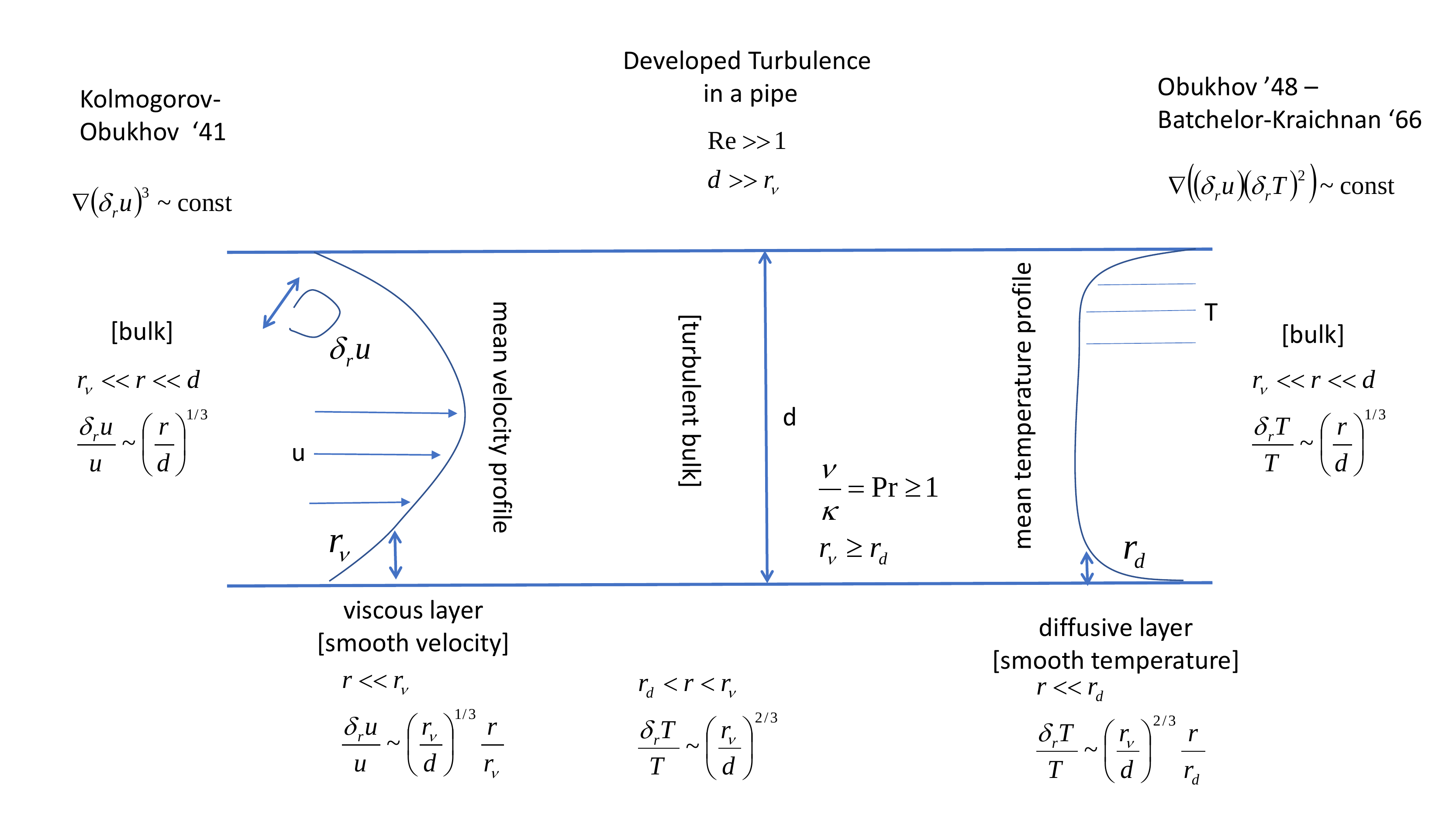}
\caption{Schematic illustration of boundary layers' layout and RKOK turbulent phenomenology estimations for mass-heat turbulent flows in a pipe. See \cite{95Fri} for details and extensive references. \label{fig:turb_sketch}}
\end{figure}

Basic microscopic equations governing dynamics of incompressible velocity, $u$, and temperature, $\theta$, at a moment of time $t$ and at a spatial position, $r$, within the pipe (thus in three dimensions) are (schematically)
\begin{eqnarray}
&& \rho(\partial_t u + (u\nabla) u)-\nabla p=\nu\nabla^2 u =\mbox{flow driving},\quad
\nabla u=0\label{NS} \\
&& \partial_t \theta+(u\nabla)\theta=\kappa\nabla^2 \theta=\mbox{heat driving}.\label{T-micro-eqs}
\end{eqnarray}
The equations should be complemented by respective boundary conditions. Averaging these equations over turbulent fluctuations in the pipe results
in the steady mass flow Eqs.~(\ref{Flow1},\ref{Flow2}) and coarse-grained thermo-transport Eqs.~(\ref{heat}) for the average (over cross-section) velocity, $V$, and temperature, $T$, thus dependent (potentially) only on the position along the pipe.

To estimate eddy-diffusivity and heat loss coefficients entering the coarse-grained description of Eq.~(\ref{heat}) one needs to discuss underlying picture of turbulent boundary layers, schematically illustrated/summarized in Fig.~(\ref{fig:turb_sketch}). There are actually two boundary layers, one related to velocity -- so-called viscous boundary layer -- the flow is laminar within the layer, and the other called diffusive boundary layer -- within which thermal diffusion dominates turbulent diffusion/advection. Widths of the two boundary layers are estimated within the standard Kolmogorov-Obukhov (KO) phenomenology \cite{95Fri} as follows.

Velocity self advection with viscosity and assuming that velocity fluctuations at the scale, $r$, from within the range of scales bounded from above by $d$ and from below by $r_\nu$ scales as $\delta v_r \sim V (r/d)^{1/3}$) and thus
 \begin{equation}
 r_\nu=\nu^{3/4}d^{1/4}V^{-3/4}.
 \label{heat3}
 \end{equation}
We assume that $\nu>\kappa$ (in water the so-called Prandtl number, defined as $Pr=\nu/\kappa$, ranges from $13$ at $0C$ to $\sim 2$ at $100C$.) Then the thermal diffusion boundary layer is found within the larger viscous boundary layer.  Diffusion dominates close to the walls, i.e. within the diffusion layer, while advection takes over on the interface with the bulk. Estimation for the width of this layer, $r_d$, from the Obukhov-Batchelor-Kraichnan (OBK) theory of scalar transport in turbulence \cite{95Fri} (assuming that velocity fluctuations in the range of scales bounded from above by $r_\nu$ and by $r_d$ is $\delta v_r\sim V (r_\nu/R)^{1/3} (r/r_\nu)$ and balancing turbulent advection and thermo-diffusion term at $r_d$) is
 \begin{equation}
 r_d=\kappa^{1/4}\nu^{1/2}d^{1/4}V^{-3/4}.
 \label{heat3}
 \end{equation}
The general picture of the microscopic temperature profile, $\theta$, across the cross-section of the pipe is roughly as follows. $\theta$ is constant in  the central portion of the pipe and then it drops to zero  in the thermo-diffusive layer, where thermal flux (losses to the wall) balances thermal diffusion. Therefore, balancing integral of the two terms  over circumference of the pipe one estimates, that $\gamma d\sim \kappa/r_d$, resulting in
Eq.~(\ref{eddy_loss}).

This concludes our brief recount of the KO-OBK phenomenology as applied to turbulent incompressible flows and scalar transport in a pipe. (See \cite{95Fri} and references there in for more details.)  In the following Subsection we present estimation for the thermo-flow characteristics  typical for pipe flows in DHS.

\subsection{Effective Parameters for an Exemplary Pipe}
\label{subsec:exemp}

{\small
\begin{center}
\begin{tabular}{|c|c|c|c|c|c|c|c|c|c|}
\hline
from & $L$, m & $d$, m & $V, \frac{m}{s}$ & $D, \frac{m^2}{s}$%\sim V R$
& $Re$
& $\bar{t}$
, s & $\Delta L$
, m & $\Delta t$
, s & $\gamma, s^{-1}$\\
\hline
\cite{08NN} & $200$  & $0.4$  & $0.8$  & $0.16$ & $3.2\cdot 10^5$ & $250$ & $6.3$ & $8$ & 0.08\\
\hline
\cite{06GBLS} & $800$  & { $0.2$}  & $0.04$  & { $0.004$} & { $8\cdot 10^4$} & $2\cdot 10^4$ & { $9$} & { $220$} & { 0.02}\\
\hline
\end{tabular}
\label{table:cases}
\end{center}
}

Parameters corresponding to exemplary cases from Russia \cite{08NN}, and Denmark \cite{06GBLS} are shown in the Table \ref{table:cases}. Here, $L$ is the length of the pipe/system, $d$ is diameter of the pipe, $V$ is the flow velocity, $D\sim V d$ is the eddy/turbulent diffusion coefficient, $Re=D/\nu$ is the Reynolds number where $\nu\approx 5\cdot 10^{-7} m^2/s$ is the kinematic viscosity of the water (at 60C), $\bar{t}=L/V$ is the expected time of propagation (e.g. of the heat front) through the system, $\Delta L=\sqrt{D \bar{t}}=\sqrt{d L}$ uncertainty in the front position due to eddy diffusivity, $\Delta t=\Delta L/V=\sqrt{d L}/V$ uncertainty in the time of propagation through the system due to turbulence, $\gamma $ is turbulent thermal loss coefficient, estimated according to Eq.~(\ref{eddy_loss}, and one takes $\kappa\approx 1.8\cdot 10^{-7}\ m^2/s$ for water (at 60C).

\subsection{Heat flux}
\label{subsec:heat_flux}

An important characteristic of interest is the heat flux transferred along the pipe at the position, $x$, and the moment of time, $t$:
\begin{eqnarray}
&& q(t;x)=\frac{Q(t;x)}{\rho c_p \pi (d/2)^2} \doteq  V T-D\partial_x T,
\label{heat_flux}\\
&& \hspace{-1cm} \rho c_p \pi (d/2)^2
\approx 10^3 \frac{kg}{m^3}\cdot 4.2\cdot10^3\frac{W\cdot s}{kg\cdot K}\cdot\pi (0.25\cdot m)^2
\approx 8.2\cdot 10^6\frac{W\cdot s}{K\cdot m},\label{heat_water}
\end{eqnarray}
where the two terms on the right hand side of Eq.~(\ref{heat_flux}) correspond to the advective contribution (which may also be called mass flow or hydraulic contribution) and (turbulent, eddy-) diffusive or convective contribution; $\rho c_p \pi (R/2)^2$ sets the units for the heat flux, so that $Q$, representing power,  is measured in [Watts], while $q$ is measured in [m K/s]; $\rho$ is the carrier/fluid density, $c_p$ is the heat capacity of the carrier/fluid, and $\pi R^2/4$ is the area of the pipe cross-section;  Eq.~(\ref{heat_water}) gives an exemplary estimate for water. Notice, that the  split on advective/hydraulic vs diffusive/conductive terms is formal in the turbulent regime where, according to Eq.~(\ref{eddy_diff}), the two mechanisms originate from the same physics and cannot be separated. Notice that in the normal operational regime heat flow goes along the mass flow and also against the temperature gradient, i.e. the two terms on the rhs of Eq.~(\ref{heat_flux}) work in unison, even though one expects that the first advective/hydraulic term is significantly larger, since $L\gg R$. (In principle the small advective correction may be of the opposite sign in the system where water in the pipes is kept at the temperature lower than the ambient/outside temperature. However,  these cooling regimes are exotic, if realistic at all.)

\section{Cauchi problem: formulation and solution}
\label{sec:Cauchi}

{ In this Section we discuss} solution of the thermal transport Eq.~(\ref{heat}) describing propagation of a thermal front through a district heating pipe.

\begin{figure}[t]
\centering
\includegraphics[width=0.33\textwidth]{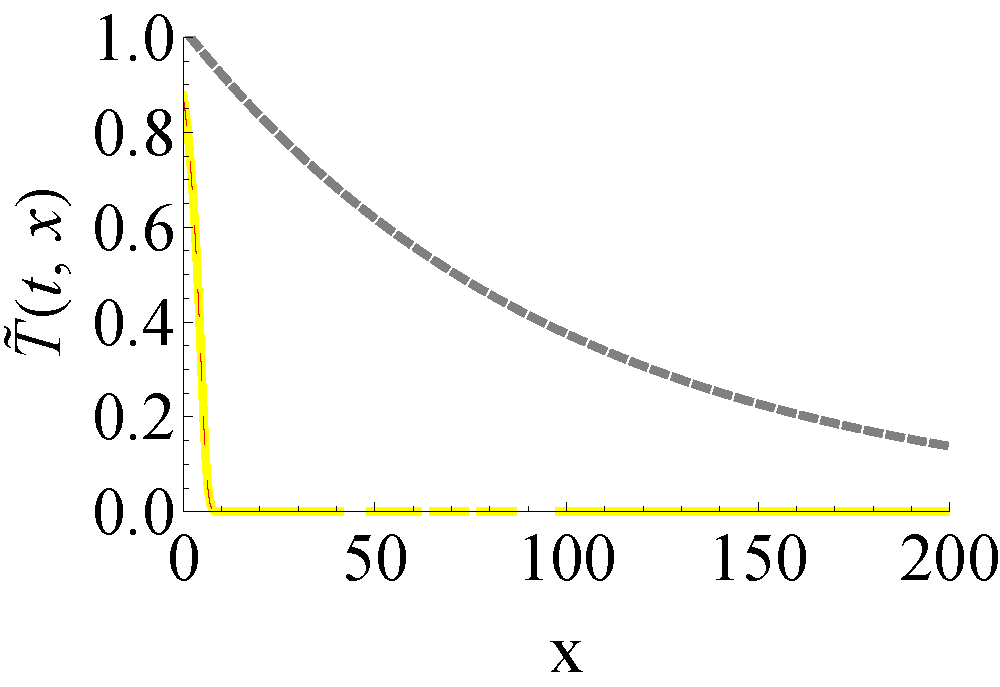}
\includegraphics[width=0.33\textwidth]{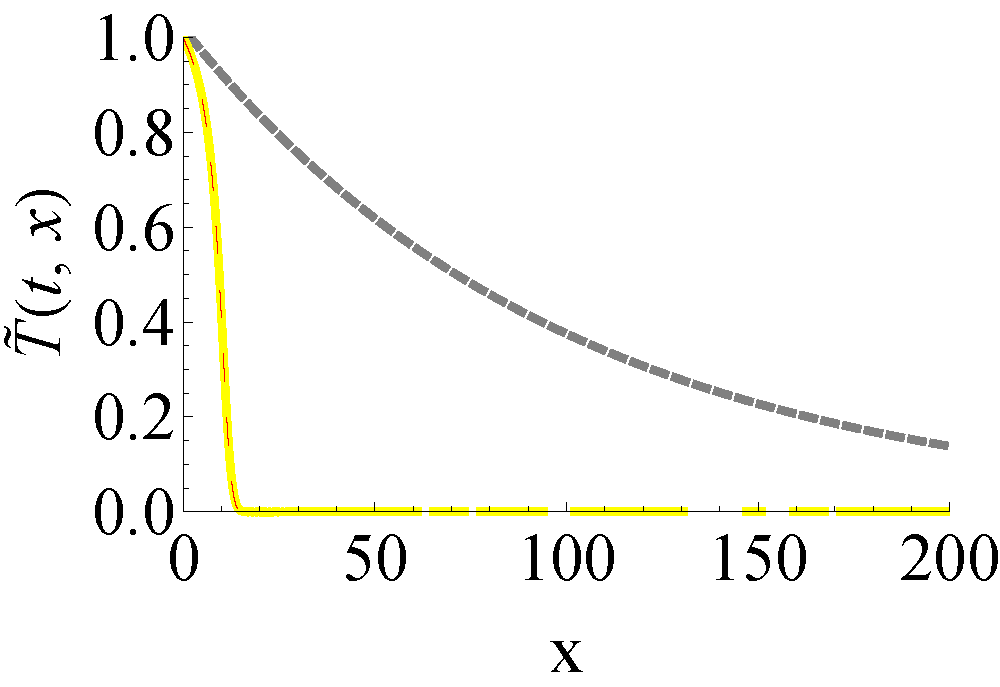}
\includegraphics[width=0.33\textwidth]{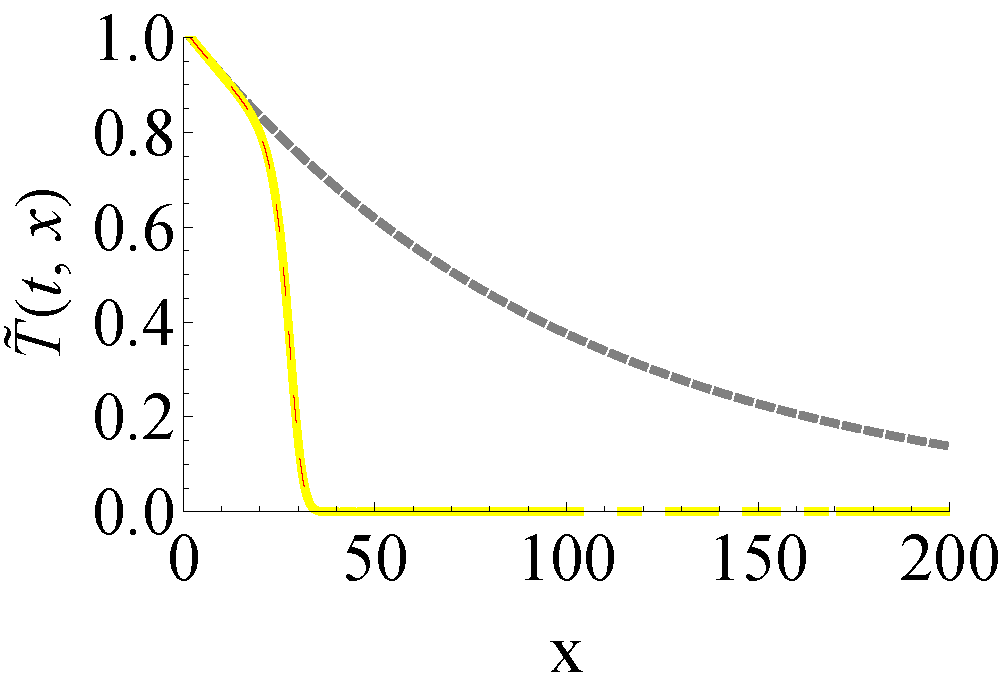}
\includegraphics[width=0.33\textwidth]{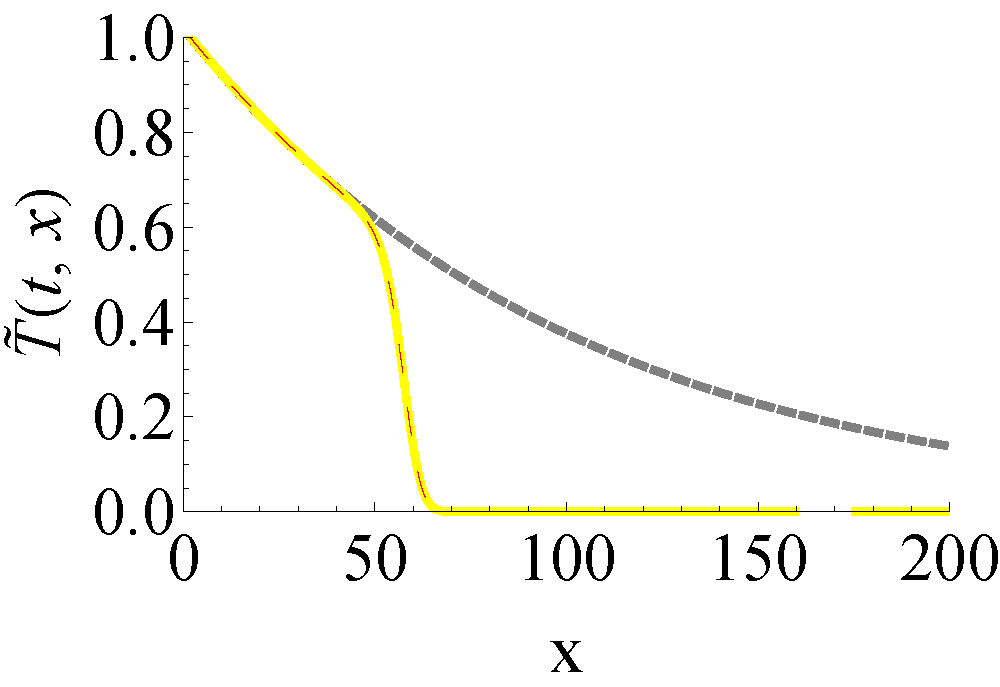}
\includegraphics[width=0.33\textwidth]{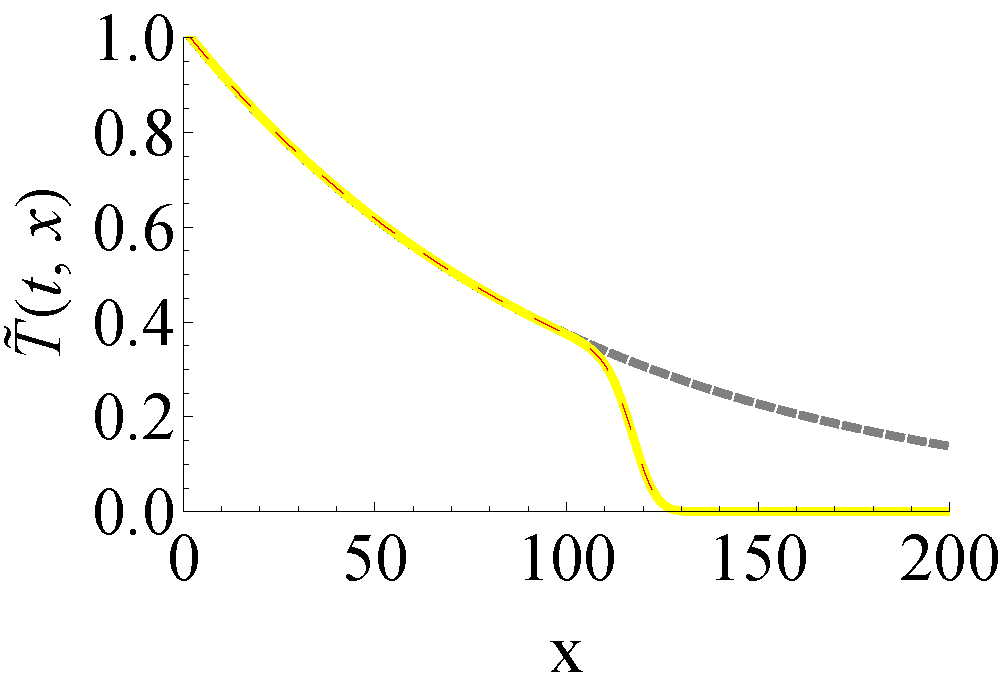}
\includegraphics[width=0.33\textwidth]{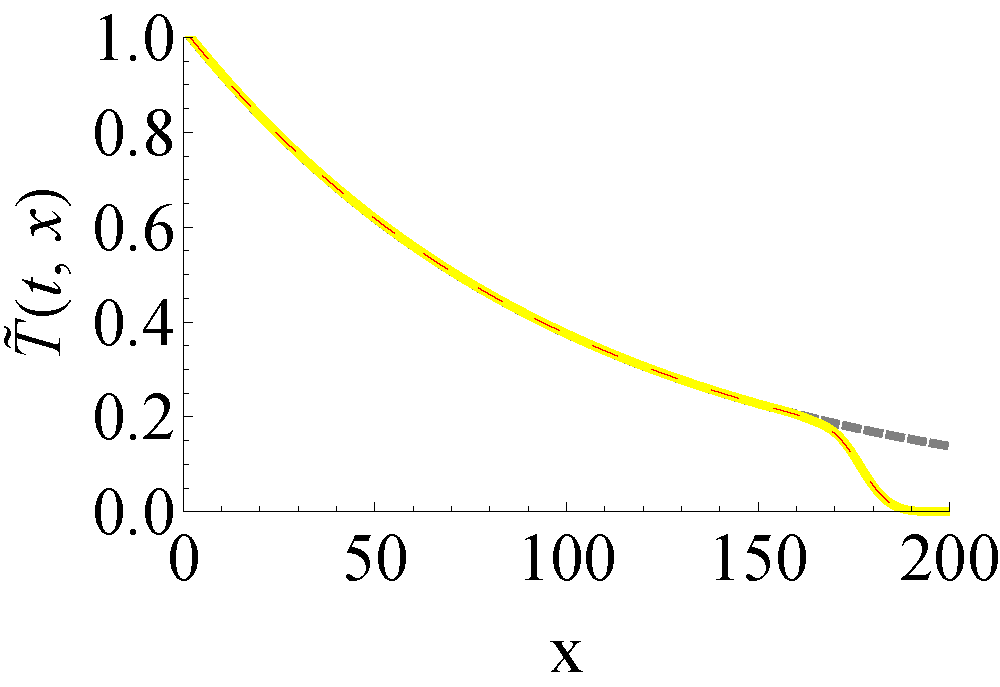}
\caption{Diffusive spread of the front.
$T_{st}(x)$ is a stationary (time independent) profile (shown red).
$T^{(in)}(t)=T_{st}(0)+\Delta \left(1-\exp(-t/\tau)\right)$
$\tilde{T}(t,x)=T(t,x)-T_{st}(x)$ is shown.
$V=0.1, L=100, \gamma=0.002, D=0.08, \tau=10$, $t=[0.01,0.02,0.05,0.1,0.2,0.3]$ [SI units].
Gray dashed shows asymptotic stationary solution. Solid yellow shows numerical solution of PDE. Dashed red shows explicit expression in terms of a convolution with the (advection-diffusion) kernel.
Noticeable features: (a) front propagates with the flow;
(b) Diffusive Spread is significant dynamically, however its effect is suppressed by (a small parameter) $R/L$ in the steady state; (c) minor bump trailing behind the front propagates with the same velocity (as the front).\label{fig:Cauchi_dynamics}}
\end{figure}

Without loss of generality we assume that initially, i.e. at $t=0$, temperature profile of the temperature in a single pipe of length $L$ is
the steady time-independent solution of Eq.~(\ref{heat}) correspondent to injection into the pipe of the flux $q_0$ at the inlet, $x=0$,
\begin{eqnarray}
&& \forall x\in[0,L]:\nonumber\\
&& T(0,x)=T_{st}(x)\doteq \frac{2 q_0}{V+\sqrt{V^2+4D\gamma}}\exp\left(\frac{V-\sqrt{V^2 +4 D \gamma}}{2 D}x\right).
\label{initial}
\end{eqnarray}
{ Given similarity to the so-called Shukhov equation} \footnote{
Shukhov is Russian engineer-polymath, scientist and architect \cite{Shukhov} who has suggested to model temperature decay in steady pipe flows due to heat losses via an exponential function of the distance (along the pipe). General discussion of the Shukhov formula for dependence of the temperature drop on other characteristics and properties of the flow can be found in \cite{85AKU,82Kri}. See also \cite{75Sok,82Ion,86Zin,88BVG} for extensive discussions, derivations and validations of the Shukhov formula in the regimes of interest for water-based district heating systems.}  { we call Eq.~(\ref{initial})  the generalized Shukhov equation. (To derive the original Shukhov equation from the generalized Shukhov equation, one should simply replace $D$ in Eq.~(\ref{initial}) by zero.)}

Denoting
\begin{eqnarray}
\tilde{T}(t,x)=T(t,x)-T_{st}(x).
\label{tilde_T}
\end{eqnarray}
deviations from the steady profile driven by the action of the additional (to $q_0$) heat source, $q(t)$, switched on at $t=0$ and growing to a constant, one arrives at the following equations
\begin{equation}
 \partial_t \tilde{T}+V\partial_x \tilde{T}
 -D \partial_x^2 \tilde{T} +\gamma \tilde{T}=q(t)\delta(x).
 \label{heat_source}
 \end{equation}
We recap that this setting  corresponds to enforcing  { for $\tilde{T}$ the flux Boundary Condition (BC) at the heat injection inlet of the pipe}
\footnote{ Physical consideration described above justifies this choice of the constant flux condition at the origin. Notice  that imposing other BC at the origin, e.g. so-called Neumann BC correspondent to maintaining the temperature constant at the inlet (and not the flux) may also be realistic for some, less common, heat injection sources. Even more generally a mixed BC, maintaining a linear combination of temperature and of the heat flux is also a possibility. Comprehensive discussion of how to model the heat source condition goes beyond the scope of the paper. The authors believe that this important question should be resolved in the future through an accurate model calibration against experimental/field data.},
{
\begin{equation}
\forall t>0:\quad V \tilde{T}(t,0)- D\partial_x \tilde{T}(t,0)=q(t),
\label{BC_flux}
\end{equation}
zero asymptotic condition at $x={ +}\infty$,
\begin{equation}
\tilde{T}(t,+\infty)=0, \label{BC_infty}
\end{equation}
and zero initial condition
\begin{equation}
\forall x\geq 0:\quad \tilde{T}(0,x)=0. \label{IC}
\end{equation}
}

Explicit solution of Eq.~(\ref{heat_source},\ref{BC_flux},\ref{BC_infty},\ref{IC}) becomes
 \begin{eqnarray}
 \tilde{T}(t,x)=\int\limits_0^t d t' q(t') \frac{\exp\left(-\gamma(t-t')-\frac{(x-V (t-t'))^2}{4D(t-t')}\right)}{\sqrt{4\pi D(t-t')}}.
 \label{T_source}
 \end{eqnarray}
Advancement of the heat front, its diffusive spread and the follow up transient past the front settling into a steady profile are illustrated in Fig.~(\ref{fig:Cauchi_dynamics}). Discussion  of other possible choices of the { BC (see the footnote)}, e.g. corespondent to maintaining the temperature constant at the { inlet, is discussed} in \ref{app:BC_temp}.

\section{Temperature (thermal) advection-diffusion equations in a multi-pipe system (network)}
\label{sec:network}

Consider stationary flows and, moreover, ignore dependence of the flows on density and temperature,  assuming that changes in the latter
are sufficiently small to have a significant thermodynamic effect on density, $\rho$, considered known and constant through out the multi-pipe system at the time scale of consideration (through out few hours). Then, the system of Eqs.~(\ref{Flow1},\ref{Flow2}) are separated (from dynamics of the temperature field and heat fluxes) and shall be re-solved first. Output of the hydro-static analysis is the set of velocity values and mass flows associated with each edge (pipe) of the system
\begin{eqnarray}
\forall \alpha\in {\cal E}:\quad V_\alpha=\frac{\phi_\alpha}{\pi (d_\alpha/2)^2\rho}
\label{v-network}
\end{eqnarray}
thus found and assumed known. Here in Eq.~(\ref{v-network}) $d_\alpha$ is the diameter of the pipe $\alpha$ and $\rho$ is the density of fluid/water. One also assumes that $\alpha$ marks a label of a directed edge oriented along the direction of the flow of the given pipe and that the diameter of the pipe is constant (does not change along the pipe), and thus the velocity of the incompressible flow is constant (along the pipe) too.

The hydro-static description of flows by Eq.~(\ref{Flow1},\ref{Flow2},\ref{v-network}) is to be complemented by the advection-diffusion equations discussed in the following Subsection.

\subsection{Formulating and Solving Stationary Advection-Diffusion-Losses Equations over the Network}
\label{subsec:network_static}

\begin{figure}[t]
\centering
\includegraphics[width=0.9\textwidth]{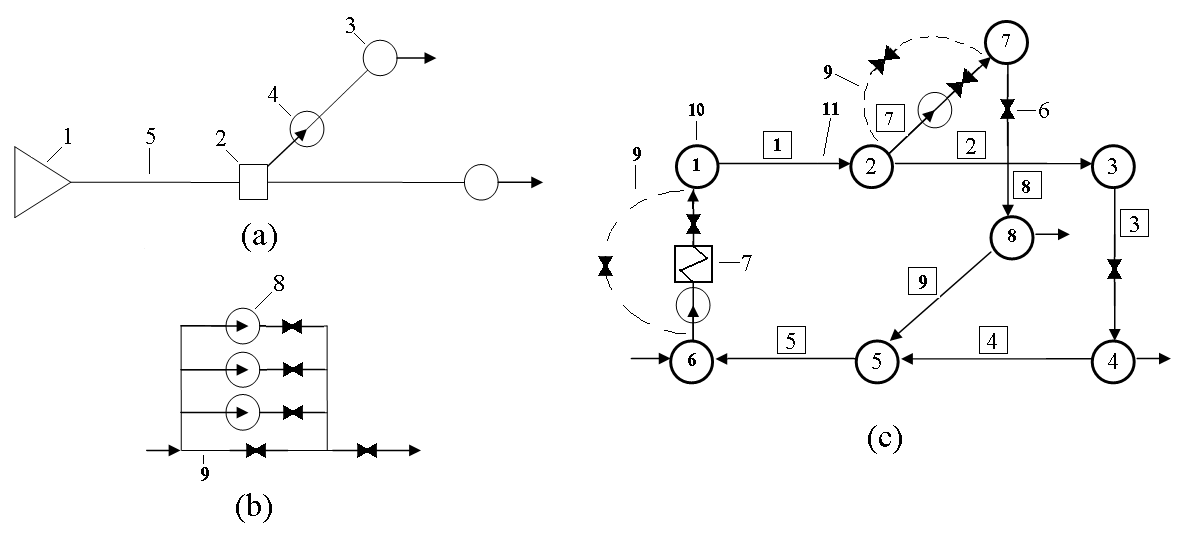}
\caption{Example of a district heating technological scheme (a) pump station; (b) two-pipe (hot and cold combined) computational scheme (respective hot/heat-delivering and cold/carrier-returning pipes are shown together - as one); (c) 1- heat source, 2- heat chamber, 3- consumer, 4- pump station, 5- pipe, 6-control unit, 7- heaters, 8 - pumps, 9 - bypass pipe, 10 - nodes (denoted as $a\in{\cal V}$ below), 11- pipe/ (undirected) edge (denoted as $\{a,b\}$ below, note that our notation for the directed edge/pipe is $(a,b)$).
\label{fig:techno}}
\end{figure}

\begin{figure}[t]
\centering
\includegraphics[width=0.4\textwidth,page=1]{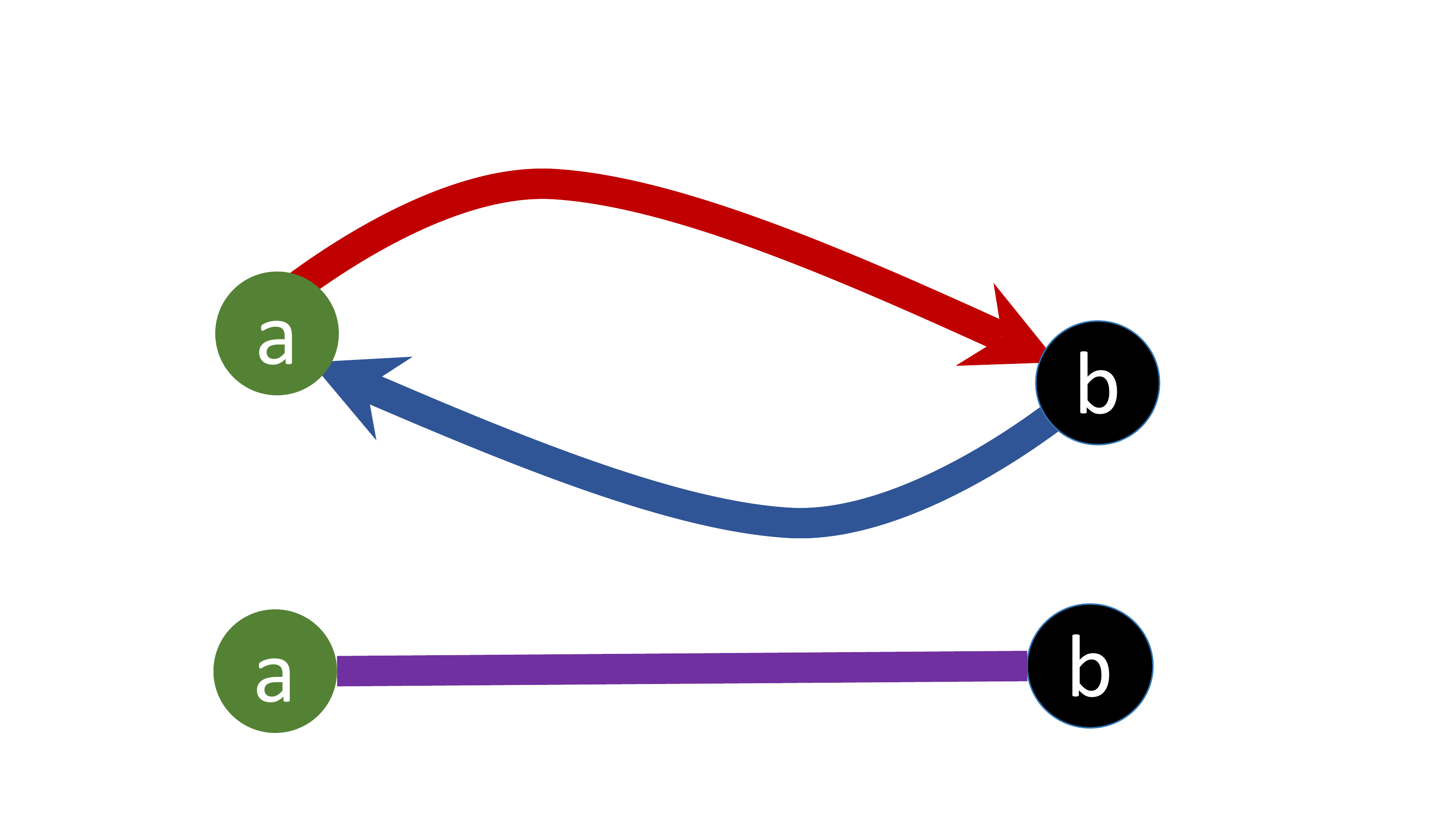}
\includegraphics[width=0.4\textwidth,page=2]{Figs/exemplary_network.pdf}
\caption{Schematic layout of an illustrative district heating network with two (left) and three (right) active heat loads (consumers of heat marked black and producer of heat marked green). In both cases the two-pipe (hot and cold combined) and single-pipe (hot and cold separately) schemes are shown on the top and bottom parts of the plots respectively. For the three active node example there are also two branch/mixing nodes, one branching node for the hot part (heat delivery) of the network, colored red, and one branching node for the cold part of the network (return of the cold carrier to the heat source).  The hot and cold parts of the network are colored red/blue in the two-pipe schemes and respective (combined) lines are shown purple in the single-pipe schemes.  Arrows in the two-linear schemes show direction of the flow. More details, e.g. on description of temperature mixing model at a branching node and on dependence of the temperature drop at the source/consumers on the local amount of heat produced/consumed are discussed in the text. \label{fig:exemplary_network}}
\end{figure}

Consider a thermo-static case when all the temperatures and heat flows are stationary. Temperature field distribution over a pipe for given boundary conditions on temperature (or heat flux) was described above, thus what remains to be done is to complement this ``along the pipe" description with providing details on the temperature change at other ``nodal" elements of the system.

Layout of an actual/technological system is rather involved, see e.g. Fig.~(\ref{fig:techno}) showing the base elements and an exemplary computational scheme. However,  aiming to describe only principal modeling we will not attempt to reproduce all the practical details here, instead we will focus on a somehow simplified view of a district heating network containing only heat branching/mixing nodes and producing/consuming nodes, thus skipping to describe such practically important elements of the district heating system as pumps, heating chambers, etc. In the following we will work primarily with the single-pipe representation where hot and cold pipes are considered separately. See e.g. top sub-figures/portions in Fig.~\ref{fig:exemplary_network}.

In fact, aiming to have a simple modular description we will not differentiate between branching nodes and active nodes, combining these in a generalized node, $a\in {\cal V}$, each represented by a set of incoming pipes, $\alpha \in \mbox{in}(a)\subseteq {\cal E}$, and outgoing pipes, $\alpha \in \mbox{out}(a)\subseteq {\cal E}$, where ${\cal E}$ represents the set of directed (oriented along the flow) edges/pipes, and ${\cal V}$ represents the set of nodes. Then heat fluxes of the pipes outgoing a generalized node, $a$, are expressed via incoming heat fluxes, $q_{a;in}=(q_{\alpha\to a}|\alpha \in \mbox{in}(a))$, and of the production/consumption of heat, $Q_a$, at the node
\begin{equation}
\label{node-relations}
\forall a\in {\cal V}:\quad \forall \alpha\in \mbox{out}(a):\quad
q_{a\to\alpha}=G_{a\to \alpha}(q_{a;in}; Q_a). \label{G-function}
\end{equation}
The $G$-functions, representing models of the nodes, can be rather different for different nodes. Here we present some most popular exemplary models. A perfect mixing model of the branching node without injection/consumption of heat, i.e. with $Q_a=0$, is
\begin{eqnarray}
& \underline{\mbox{Branching node with perfect mixing}}:& \nonumber\\
&& \hspace{-4cm} G_{a\to \alpha}(q_{a;in}; Q_a=0)=\frac{\pi \rho V_\alpha}{4}\frac{
\sum_{\beta\in\mbox{in}(a)} q_{\beta\to a} (d_\beta)^2}{\sum_{\gamma\in\mbox{in}(a)} \phi_\gamma}.
\label{branching_mixing_q}
\end{eqnarray}
A standard model of a production/consumption node, $a$, with one input pipe, $\mbox{in}(a)=\alpha$ and one output pipe, $\mbox{out}(a)=\beta$ is
\begin{eqnarray}
& \underline{\mbox{Standard production/consumption node (single input, single output)}}:& \nonumber \\
&& \hspace{-10cm} G_{a\to \alpha}(q_{\beta\to a}; Q_a)=\frac{\rho c_p\pi (d_\beta/2)^2q_{\beta \to a}+Q_a}{\rho c_p\pi (d_\alpha/2)^2},
\label{standard_production_consumption_q}
\end{eqnarray}
where $Q_a$ is heat injected (then $Q_a>0$) or consumed (then $Q_a<0$) at the node $a$.

According to Eq.~(\ref{initial}), which is the stationary solution of the basic Eq.~(\ref{heat}), heat flux at the outlet of the pipe, as expressed via the heat flux at the inlet, is
\begin{eqnarray}
 &&
\underline{\forall \alpha=(a,b)\in{\cal E}}: \nonumber\\
&& q_{\alpha\to b}=q_{a\to\alpha}\exp\left(\frac{V_{\alpha}-\sqrt{V^2_{\alpha} +4 D_{\alpha} \gamma_{\alpha}}}{2 D_\alpha}L_{\alpha}\right), \label{Talpha_b}
\end{eqnarray}
where (as before) the in- and out- sides of the pipe are defined with respect to direction of the mass flow.

Eqs.~(\ref{G-function}), where the respective G-functions can be chosen, for example, according to Eqs.~(\ref{branching_mixing_q}) and Eqs.~(\ref{standard_production_consumption_q}),
supplemented by Eqs.~(\ref{Talpha_b}) and defined over a single-connected district heating network,
set up an example of the steady ``network heat flow" problem:
\begin{itemize}
\item \underline{\bf Input:}
\begin{itemize}
\item Heat flows at all the pipes leaving all the production nodes of the network;
\item Powers consumed at all the consumer nodes;
\end{itemize}
\item \underline{\bf Output:}
\begin{itemize}
\item Heat flows at all other locations (not given within description of the input);
\item Temperatures at all network locations;
\item Powers injected at the production nodes.
\end{itemize}
\end{itemize}
Consistently with what was discussed in the literature, see e.g. \cite{15MN}, this  set up  has a unique solution. Indeed, to verify this we need to start from generators and solve the set of Eqs.~(\ref{G-function},\ref{branching_mixing_q},\ref{standard_production_consumption_q},\ref{Talpha_b}) recursively, one by one, simply following the directed (in the single-pipe representation) graph of the district heating network. Notice that since the system of equations is linear,  re-scaling the solution (one which was just found propagating relations along the directed graph) by multiplying all heat flows, power injections/consumptions and temperature by the same positive constant number will generate another valid solution of the system of Eqs.~(\ref{G-function},\ref{branching_mixing_q},\ref{standard_production_consumption_q},\ref{Talpha_b}). This one-parametric degree of freedom can be useful for finding solutions of the system of equations in the settings where boundary conditions (input characteristics in the description above) arrange differently. For example, it may be more practical to consider a setting where heat flows at all the pipes leaving all the generators of the network are replaced by powers injected at all the generators of the network. (These two different sets of input parameters are in the one-to-one relation.) Another option for a valid input set is to replace some (or all) of the injected/consumed powers at the producers/consumers by temperatures - one per node (say outgoing for producer and incoming for consumer).

\subsection{Formulating and Solving Dynamic/Transient Advection-Diffusion-Losses Equations over Network}
\label{subsec:network_dynamic}

Generalization of the static version of the "heat flow" problem posed in the preceding Subsection to the dynamic/transient case is straightforward - one just needs to substitute Eqs.~(\ref{Talpha_b}) by the following solution of the dynamic equations derived from Eq.~(\ref{T_source})
\begin{eqnarray}
&&\underline{\forall \alpha=(a,b)\in{\cal E}}:\quad q_{\alpha\to b}(t)=  \label{Talpha_b_dyn}\\
&& \int\limits_0^t d t' q_{a\to\alpha}(t') \left(V_\alpha+\frac{L_\alpha}{2(t-t')}\right)\frac{\exp\left(-\gamma_{\alpha}(t-t')-\frac{(L_{\alpha}-V_{\alpha} (t-t'))^2}{4D_{\alpha}(t-t')}\right)}{\sqrt{4\pi D_{\alpha}(t-t')}}.
 \nonumber
\end{eqnarray}
Let us emphasize that explicit form of Eq.~(\ref{Talpha_b_dyn}) in quadratures makes the dynamic problem as efficient computationally as the static problem.

\subsection{Dynamic Network Illustration}
\label{subsec:two_pipes}

\begin{figure}[t]
\centering
\includegraphics[width=0.8\textwidth]{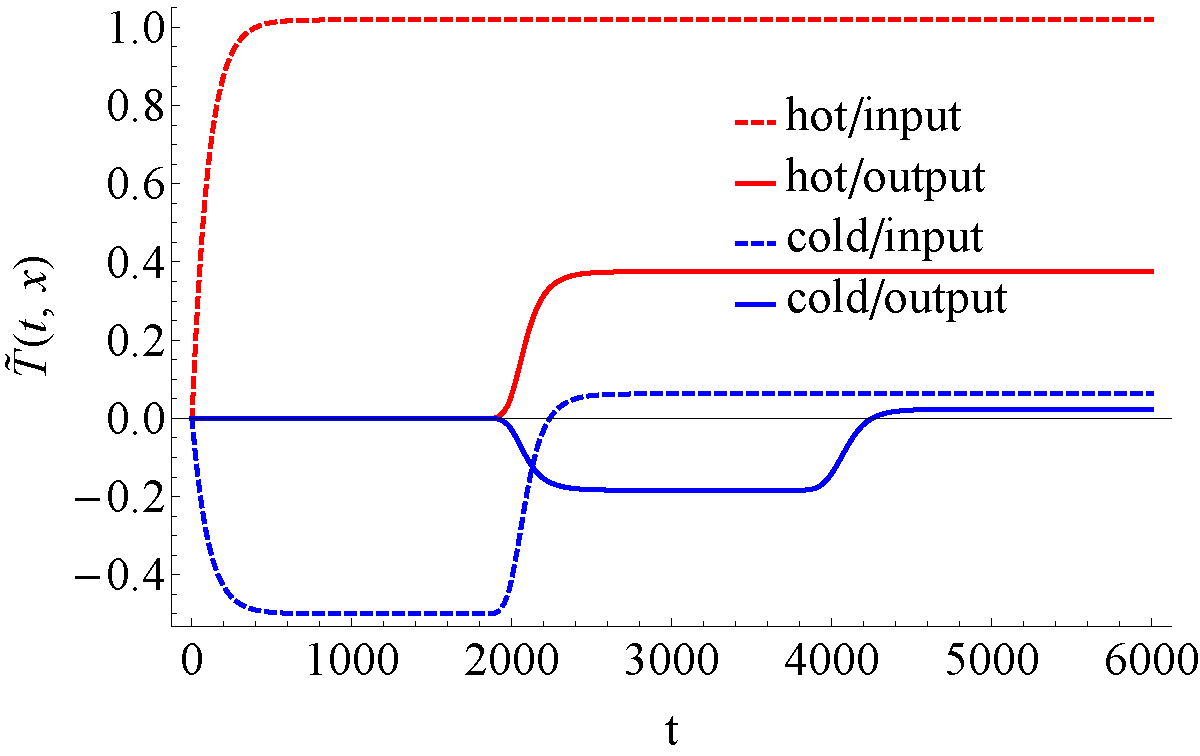}
\includegraphics[width=0.95\textwidth]{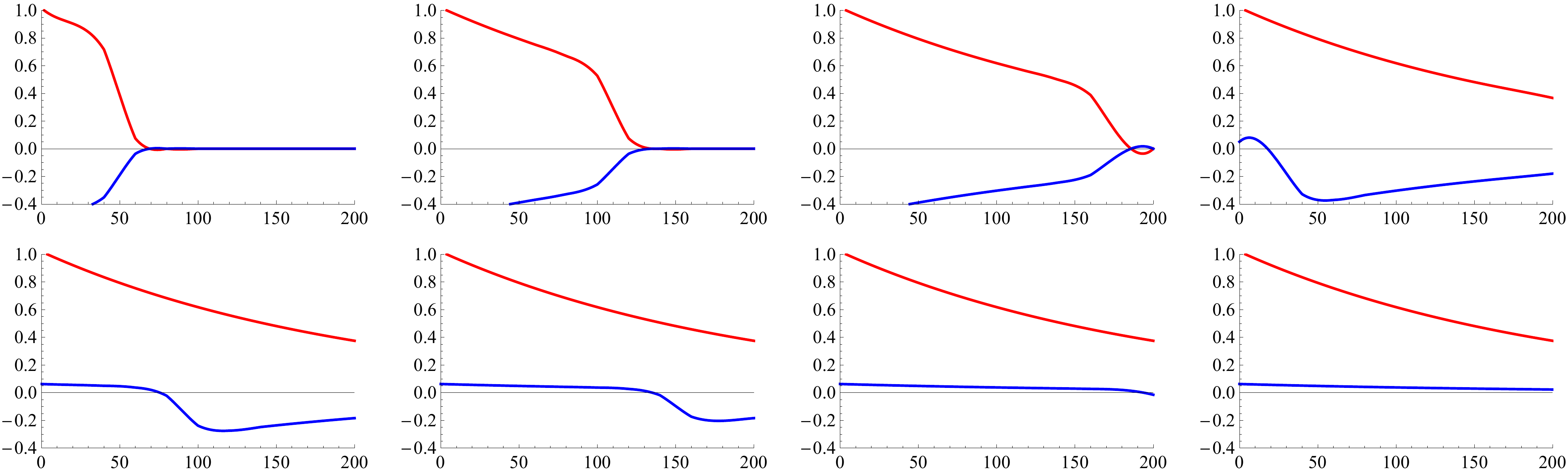}
\caption{Top sub-figure: dependence of temperature deviation (from the initial stationary profile) at the four principle locations on time for the dynamica/transient process in the two-pipe example of Fig.~\ref{fig:exemplary_network}. Bottom sub-figure: 8 consecutive snapshots showing temperatures at the output of the hot and cold pipes (red and blue respectively). Parameters of the two identical pipes are chosen equal to parameters of the single pipe from the illustration discussed in Section \ref{sec:Cauchi}). The system is driven by a prescribed time-dependent profile of the heat flux at the heat production location into the hot pipe, $q_{hot}(t;0)=q_{in}(1-\exp(-t/\tau))$, and prescribed consumption of heat at the consumer, $q_{hot}(t;L)-q_{cold}(t;0) = q_{out}(1-\exp(-t/\tau))$. We choose, $q_{in}=1.02 u$ and $q_{out}=0.5 u$, where $u$ is the (constant) velocity of the stationary flow.  \label{fig:two_pipes}}
\end{figure}

\begin{figure}[t]
\centering
\includegraphics[width=0.8\textwidth]{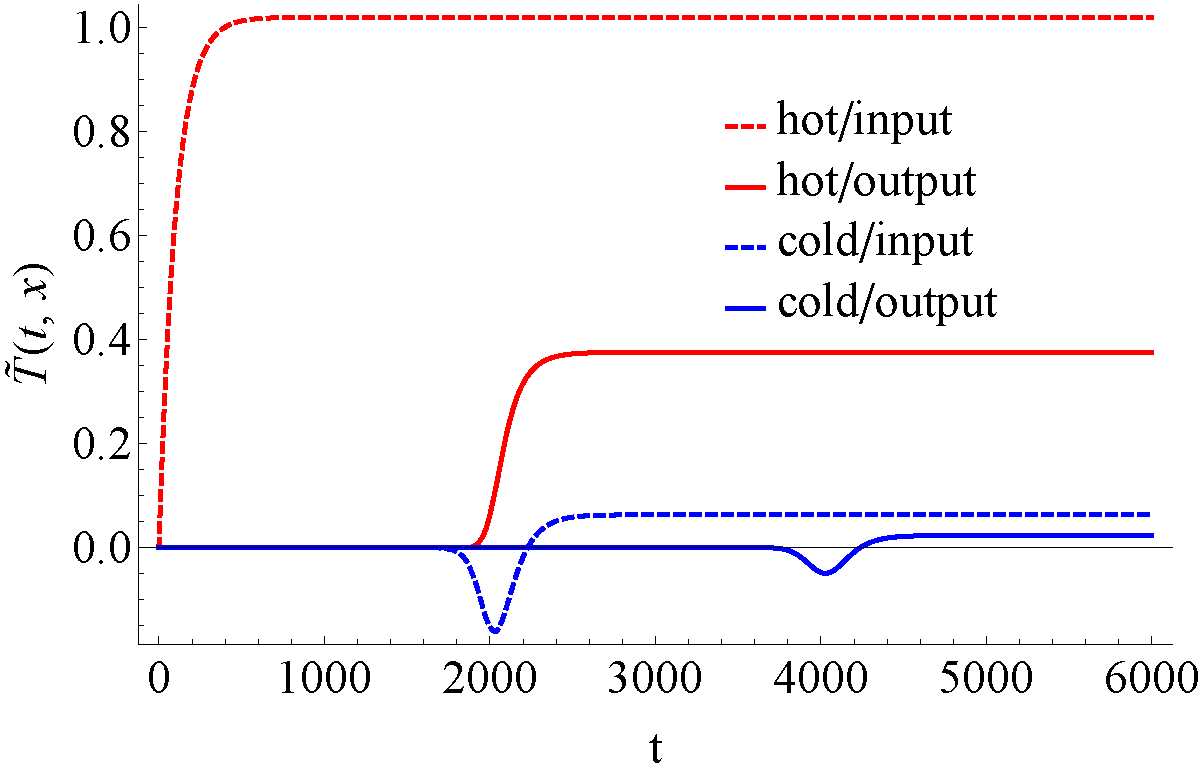}
\includegraphics[width=0.98\textwidth]{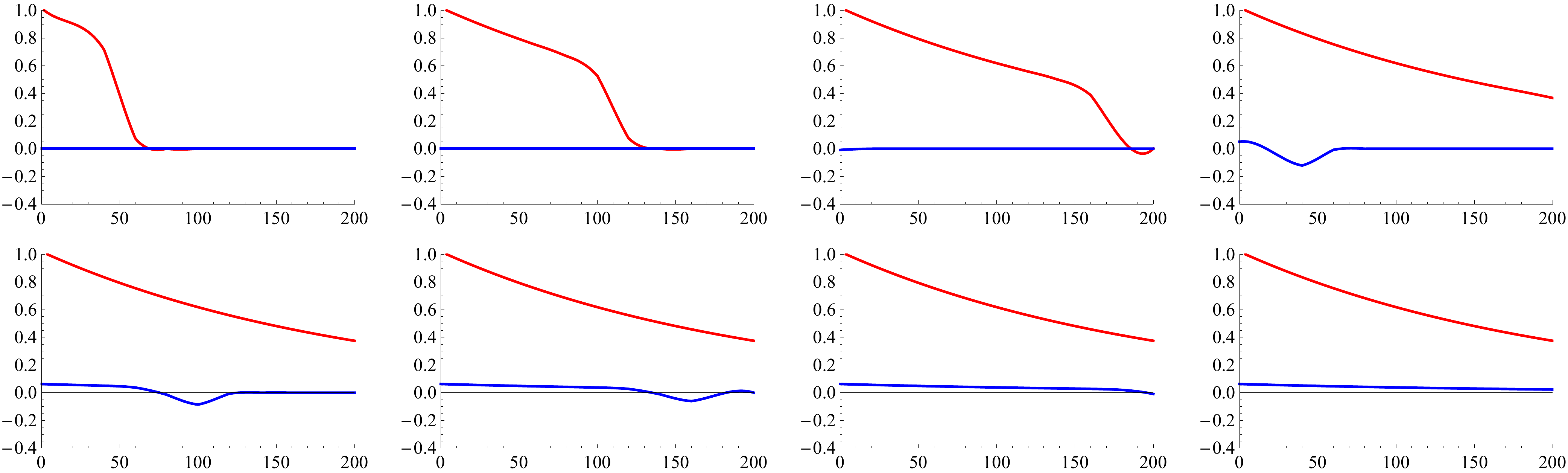}

\caption{The set of figures identical to these shown in Fig.~(\ref{fig:two_pipes}) under exception that the consumption of heat is arranged with a delay according to $q_{hot}(t;L)-q_{cold}(t;0) = q_{out}(1+\tanh(t-t_{del})/\tau)/2$, where we choose $t_{del}=2000$.\label{fig:two_pipes_delay}}
\end{figure}

We demonstrate the quality and efficiency of our dynamic/transient advection-diffusion-losses modeling on the simple example of one producer and one consumer connected by two pipes (hot and cold) shown in Fig.~\ref{fig:exemplary_network} (right). We choose parameters for the two pipes the same and correspondent to the single pipe illustration shown in Fig.~\ref{fig:Cauchi_dynamics}. The results for two different heat injection/consumption settings are shown in Figs. ~(\ref{fig:two_pipes},\ref{fig:two_pipes_delay}). In the first case injection and consumption of heat are started simultaneously,  while in the second case consumption of heat starts with a delay. Comparison of the two examples illustrates benefits and importance of the dynamic modeling accounting for time delays and details of the heat front spreading as it advances through the network.

\section{Conclusions and Path Forward}
\label{sec:conclusions}

In this manuscript we started to explore new phenomena related to the temporal dynamics in the District Heating Systems. Specifically, we have made the following set of statements and observations:
\begin{itemize}
\item Dynamics of the thermal field in DHS pipes, analyzed over distances which are longer than diameter of the pipe and at temporal scales which are longer than the sound-wave transients, is governed by the set of coupled linear advection-diffusion-heat loss equations.  Parametric dependence of the turbulent diffusion and loss coefficients on mean velocity of the turbulent flow in a pipe and microscopic characteristics of the heat carrier (water) were estimated, according to standard turbulent phenomenology, in the practical regime of large Re-numbers.

\item Solutions of the advection-diffusion-heat loss equations were analyzed under rather general initial and boundary conditions. Description of a heat front, initiated by a change of temperature or heat flux at the heat source or sink,  was in the focus of our analysis which captures principal qualitative features of the phenomena, such as diffusive spread of the front. General solution was presented in quadratures and compared with direct simulations. The quadrature representation of the general solution in the form of an integral convolution of the heat kernel with the function representing the heat source(s) is explicit.

\item We have explained, in principle and on a simple example, how the single pipe results allow efficient and accurate computations of thermal transients in a network. The problem   reduces to explicit recurrency/propagation  of solutions from the heat source to consumers and back thus bypassing solving the system of computationally expensive Partial Differential Equations (PDEs) over the network.
\end{itemize}

In relatively short terms the results presented in the manuscript call for further research and exploration of the following topics:
\begin{itemize}
\item {\it Demonstration of the power of the approach on a realistic district heating network.} { This should also include calibration of the freedom in modeling heat sources/producers and sinks/consumers.}

\item {\it Extension of the network-wide computations to variety of the operational regimes of interest.} In this manuscript we have focused solely on describing regime when the mass/velocity flows are kept steady. However,  the methodology is extendable to other regimes of interest,  e.g.  where the heat flux is regulated through control/change of the mass flow.

\item {\it Accounting for statistical effects and uncertainty.}
\end{itemize}

Further down the road we plan to utilize the approach and address more complex problems, such as
\begin{itemize}
\item {\it Data driven validation of the network parameters and robust modeling.} District heating systems are subject to relatively fast changes acquired through operational wear and tear and related maintenance services. This is typically seen through a rather significant change of the key elements of pipes' and devices' parameters acquired in the time frame of a month (especially during the active/cold season). Thus, effective diffusion coefficients and effective heat-loss coefficients, influencing the thermal dynamics in the network, will change on this time scale and may be uncertain.  There are two approaches we plan to develop to account for the complications.  First of all,  we will be developing efficient data driven approaches which allow to reconstruct parameters from the minimal amount of field measurements. Second,  even if the data driven reconstruction is employed one still do not expect to reduce the parameter uncertainty completely.  Then we will rely on the probabilistic techniques to describe how uncertainties within the model parameters affects description of the heat transfer through the network.

\item {\it Development of controls for improved efficiency and security.}  Scalability of the thermal transient computations which we started to develop in this manuscript should allow to explore variety of available controls efficiently. In addition to developing simulation-based control capability, we plan to use linearity of the underlying equations and the explicit solvability property to develop efficient solutions extending classic Linear-Quadratic-Gaussian (LQG)-control approaches accounting for fluctuations. We will also explore respective robust formulations accounting for uncertainty in the system parameters.

\item {\it Short- and Long- term planning, aware of operations.} The main problem in planning reliably over a sufficiently long time horizon (year and beyond) is in accounting for fluctuations and uncertainty of the heat consumption/production and also accounting for evolution and uncertainty in the network parameters (e.g. wear and tear of pipes and other devices) year(s) ahead. Combined with seasonal weather variability, this uncertainty, growing with time calls for including  multiple consumption/production scenarios into a multi-level optimization designed to solve the planning problems. Moreover,  to resolve operational quality of service constraints, subject to delays and inertia, plausible scenarios need to account for  dynamics on the scale of hours to days.

\item{\it Including Demand Response (DR) elements in the district heating modeling and control.} In this manuscript we have modeled heat consumption according to the oversimplified linear relation between the consumed heat and the temperature drop across the device. New concepts of DR, originally developed for electric loads \cite{11CH,14Siano} but obviously extendable to heat loads, will require more general and flexible modeling of the heat consumption, which may be nonlinear, nonlocal in time and also changing with time depending on both exogenous an endogenous (control) signals.

\item {\it Integrated modeling and control of energy infrastructures.} The concept of smart grids, when extended to integrated energy systems at the scale of a district or a city, has a lot of potential. To implement the concept we will attempt to take advantage of relations, dependencies and inter-operability of the power, natural gas and district heating networks. See, e.g., a recent example of a conceptually new take at the inter-operability of energy infrastructures considered in \cite{15OEB}, where a unifying view on photovoltaic/natural gas power and cooling systems was proposed. To model joint operation of the three major energy infrastructures one will need to build and integrate dynamic models for all the relevant variables over the range of time scale starting from minutes and extending to days (and possibly even beyond).  The technique developed in this manuscript to model heat flows in the district heating system is expected to contribute dynamic modeling, control and planning of the future integrated energy systems.
\end{itemize}

\appendix

\section{  An alternative Boundary Condition  at the heat injection inlet of the pipe}
\label{app:BC_temp}

{ As mentioned briefly in the main text, the choice of the Neumann, i.e. constant temperature, BC at the outlet of a pipe (or a bit more accurately at the hypothetical infinite outlet) is unambitious. However, selection of the BC at the heat injection, inlet, side of the pipe reflecting the nature of the heat injection is much less obvious. We suggest, based on physics considerations, that choosing the mixed BC corresponding to zero heat flux at the inlet is appropriate for representing a perfect heat source. This is simply because a good (ideal) heat source is build with the purpose of injecting the prescribed/controled amount of heat. However to reach an univocal conclusion on if the heat source model represents reality we will need to verify it through comparison with actual real world experiment on an operational heat source. This type of experimental test is out of scope of the paper. However, to provide input in this important line of future research we have experimented (in simulations) with different type of BC on the inlet (Neumann, Dirichlet or what we believe is the correct one - heat-flux constant) and have concluded that even thought the change in the inlet BC leads to some variability in a vicinity of the inlet, also proliferating along the pipe through an overall rescaling, the main physical picture of the front propagation and spreading remains invariant with respect to the choice of the inlet BC.

Therefore, in this Appendix we discuss}, for completeness, an alternative choice of the BC at the inlet (\ref{BC_flux}), { the Neumann BC}
\begin{eqnarray}
T(t,0)=T_{st}(0)+\tilde{T}(t,0)=T^{(in)}(t),
\label{BC_temp}
\end{eqnarray}
{ and show that the change in the inlet BC preserves solvability of the single pipe dynamics, critical for scalability of the large system analysis.}

Analyzing Eqs.~(\ref{heat_source},\ref{BC_temp},\ref{BC_infty},\ref{IC}) via Laplace transform (in time) one derives
\begin{eqnarray}
&& \hat{\cal L}(s;x) T_s(x)=T_{st}(x), \label{Ts_eq}\\
&& \hat{\cal L}(s;x)\doteq s+\alpha +V\frac{d}{dx}-D\frac{d^2}{dx^2},
\label{FP_oper}\\
&&  T_s(0)=T^{(in)}_s,\quad T_s(\infty)=0,\label{Ts_bound}\\
&& T_s(x)\doteq \int_0^\infty dt e^{-t s} T(t,x),
\label{Ts}\\
&& T^{(in)}_s\doteq \int_0^\infty dt \exp(-t s) T^{(in)}(t).
\label{s_in}
\end{eqnarray}

Consider the following two, "linear"-profile and "exponential"-profile exemplary models of  $\Delta(t)\doteq T^{(in)}(t)$
\begin{eqnarray}
&& \Delta^{(lin)}(t)=\Delta_*\left\{
\begin{array}{cc}
\frac{t}{\tau},& 0\geq t\leq \tau\\
1,& \tau\leq t
\end{array}\right.
;\quad
\Delta^{(lin)}_s=\Delta_*\frac{1-e^{-\tau s}}{s^2\tau}; \label{Delta_lin}\\
&& \Delta^{(exp)}(t)=\Delta_*
\left(1-e^{-\frac{t}{\tau}}\right);\quad \Delta^{(exp)}_s=\frac{\Delta_*}{s(s\tau +1)}. \label{Delta_exp}
\end{eqnarray}
where $\tau>0$ and $\Delta_s$ is the respective Laplace transform.

It is convenient to seek for the general solution of Eqs.~(\ref{Ts_eq}) in the form
\begin{eqnarray}
T_s(x)=\tilde{T}_s(x)+\frac{T_{st}(x)}{s}, \label{tilde_T_s}
\end{eqnarray}
where thus $\tilde{T}_s(x)$ represents Laplace transform of the deviation of temperature from the initial stationary solution.

The resulting equation for $\tilde{T}_s(x)$ and respective BC become
\begin{eqnarray}
&& \hat{\cal L}(s;x) \tilde{T}_s(x)=0,\quad\tilde{T}_s(0)=\Delta_s\doteq \int_0^\infty dt e^{-t s} \Delta(t), %=\frac{\Delta}{s(s\tau+1)},
\label{Ts_tilde_one}
\end{eqnarray}
thus producing the following explicit solution
\begin{eqnarray}
\tilde{T}_s(x)=\Delta_s e^{x\frac{V-\sqrt{V^2+4D(s+\alpha)}}{2D}}.
\label{tilde_T_s_x}
\end{eqnarray}
Then the complete solution (in the space-time domain) is presented in quadratures via the inverse Laplace transform
\begin{eqnarray}
&& \tilde{T}(t,x)=\int\limits_{0^+-i\infty}^{0^++i\infty}\frac{ds}{2\pi i} e^{t s} \tilde{T}_s(x)
\label{tilde_T_s}
\end{eqnarray}

According to Eqs.~(\ref{tilde_T_s_x},\ref{tilde_T_s}), $\tilde{T}_s(x)$ is an analytic function of complex $s$ anywhere in the complex plain except of the singularities (typically poles) associated with $T_s^{(in)}$  and also the square root branch cut singularity along the $]-\infty, s_*-\alpha-V^2/(4D)]$ cut , where
\begin{equation}
s_*\doteq-\alpha-\frac{V^2}{4D}.
\label{s*}
\end{equation}

Shifting the integration contour in Eq.~(\ref{tilde_T_s}) accordingly, evaluating the pole integrals and
transforming the contour integral one arrives at the following explicit expression for the solution in quadratures (i.e. stated as an single-variable integral) in the case of exponential transient
\begin{eqnarray}
&&\tilde{T}^{(exp)}(t,x)=\Delta\exp\left(\frac{V x}{2D}\right) \Biggl(\exp\left(-x\frac{\sqrt{V^2+4D\alpha}}{2D}\right)\nonumber\\
&& -\exp\left(-\frac{t}{\tau}-x\frac{\sqrt{V^2+4D(\alpha-1/\tau)}}{2D}\right)\Biggr)+\Psi^{(exp)},
\label{tilde-T-full}\\ &&
\Psi^{(exp)}\!\doteq\!\frac{D\! e^{\!-\alpha t\!-\!\frac{V^2t}{4D}\!+\!\frac{V x}{2D}})}{\pi}\!\!\!\int\limits_{-\infty}^{\infty}\!\! q\sin(x q)
e^{-D q^2 t}\Delta_{s_*-Dq^2}dq.\label{Psi}
\end{eqnarray}
The remaining integral is of a Gaussian (diffusive) kernel type,  which may thus be useful to represent as a series. Expending $\Delta_{s_*-Dq^2}$ in the Taylor series over $(-Dq^2)$
\begin{eqnarray}
\Delta_{s_*-Dq^2}=\sum\limits_{n=0}^{+\infty} \frac{(-D q^2)^n}{n!}\Delta^{(n)}_{s_*},\quad \Delta^{(n)}_s\doteq\frac{d^n \Delta_s}{d s^n},
\label{Delta_series}
\end{eqnarray}
and evaluating the resulting integrals in Eq.~(\ref{Psi}) expicitly, one arrives at
\begin{eqnarray}
&& \Psi^{(exp)}=\frac{\exp\left(\!-\alpha t\!-\!\frac{V^2t}{4D}\!+\!\frac{V x}{2D}\right)x}{\pi\sqrt{D t^3}}
\sum_{n=0}^{+\infty}\frac{\Delta^{(n)}_{s_*}\Gamma\left(n+\frac{3}{2}\right)}{(-t)^n n!} \ _1F_1\left(n+\frac{3}{2},\frac{3}{2},-\frac{x^2}{4D t}\right)\nonumber\\
&& =\frac{2\exp\left(-\alpha t-\frac{(x-V t)^2}{4 D t}\right)}{\sqrt{\pi}}
\sum_{n=0}^{+\infty}\frac{\Delta^{(n)}_{s_*}}{n! (4t)^{n+1}}
H_{2n+1}\left(\frac{x}{2\sqrt{D t}}\right),
\label{Psi_ser}
\end{eqnarray}
where $H_n(y)$ is the Hermit polynomial and $\ _1F_1(a,b,z)$ is the Kummer confluent hypergeometric function.

In the case of linear transient one derives
\begin{eqnarray}
&&\hspace{-0.2cm}\tilde{T}^{(lin)}(t,x)\!=\!\frac{\Delta e^{\frac{V x}{2D}-x\frac{\sqrt{V^2+4D\alpha}}{2D}}}{\tau}
\!\!\left\{\!\!\!\!\begin{array}{cc}
\tau,& t\geq \tau \\
t\!-\!\frac{x}{\sqrt{V^2+4 D\alpha}},&\!\! 0\leq t\leq \tau
\end{array}
\right.
\!\!\!\!\!\!+\!\Psi^{(lin)},
\label{tilde-T-full_lin}\\ && \hspace{-0.2cm}
\Psi^{(lin)}\!\doteq\!\frac{D\Delta e^{-\alpha t\!-\!\frac{V^2t}{4D}\!+\!\frac{V x}{2D}}}{\pi\tau}\nonumber\\
&& \hspace{-0.2cm}  *\int\limits_{-\infty}^{\infty}\!\! dq \frac{q\sin(x q)\exp\left(-D q^2 t\right)}{(s_*-D q^2)^2}
\left\{\begin{array}{cc}
1-e^{-\tau(s_*-D q^2)},&t\geq \tau\\
1,& 0\leq t\leq \tau\\
\end{array}
\right.
\label{Psi_lin}
\end{eqnarray}

\bibliographystyle{elsarticle-num}
\bibliography{THR,TCL}

\begin{thebibliography}{10}
\expandafter\ifx\csname url\endcsname\relax
  \def\url#1{\texttt{#1}}\fi
\expandafter\ifx\csname urlprefix\endcsname\relax\def\urlprefix{URL }\fi
\expandafter\ifx\csname href\endcsname\relax
  \def\href#1#2{#2} \def\path#1{#1}\fi

\bibitem{SmartGridUS}
Smart {G}rid: {US D}epartment of {E}nergy,
  \url{https://energy.gov/oe/services/technology-development/smart-grid},
  accessed: 2017-01-07.

\bibitem{WhatIsSmartGrid}
What is smart grid?, \url{https://www.smartgrid.gov/the_smart_grid/}, accessed:
  2017-01-07.

\bibitem{SmartNaturalGas}
Natural {G}as in a {S}mart {E}nergy {S}ystem: {A}merican {G}as {A}ssociation,
  \url{https://www.aga.org/natural-gas-smart-energy-system}, accessed:
  2017-01-07.

\bibitem{SmartNaturalGasEurope}
Smart {G}rids in the {G}as {S}ector: {M}arcogas {R}eport at the {T}echnical
  {A}ssociation of the {E}uropean {N}atural {G}as {I}ndustry,
  \url{https://setis.ec.europa.eu/energy-research/sites/default/files/library/ERKC_%20TRS_Smart_District_HC.pdf},
  accessed: 2017-01-07.

\bibitem{14DCVK}
J.~Dehaeseleer, T.~Cayford, B.~de~Ville~de Goyet, I.~Kas, Gas grids for a smart
  energy system: Reports. energy grids, Gas for Energy 3 (2015) 22--26.

\bibitem{SmartDistrictHeatingEurope}
Smart {D}istrict {H}eating {C}ooling: {E}uropean {U}nion {R}eport 2014,
  \url{https://setis.ec.europa.eu/energy-research/sites/default/files/library/ERKC_%20TRS_Smart_District_HC.pdf},
  accessed: 2017-01-07.

\bibitem{14PLK}
J.~Persson, M.~Larsson, L.~Konsult, Innovations in {D}istrict {H}eating:
  {R}eport of the {I}nno{H}eat {P}roject, eu 2014,
  \url{http://innoheat.eu/wp-content/uploads/2014/02/Improvements-in-Heating-Systems.pdf}.

\bibitem{15GSV}
E.~Guelpa, A.~Sciacovelli, V.~Verda, Thermo-{F}luid {D}ynamic {M}odel of
  {C}omplex {D}istrict {H}eating {N}etworks for the {A}nalysis of {P}eak {L}oad
  {R}eductions in the {T}hermal {P}lants, ASME 2015 International Mechanical
  Engineering Congress and Exposition 6A.

\bibitem{10LMMD}
H.~Lund, B.~Moller, B.~Mathiesen, A.~Dyrelund,
  \href{http://www.sciencedirect.com/science/article/pii/S036054420900512X}{The
  role of district heating in future renewable energy systems}, Energy 35~(3)
  (2010) 1381 -- 1390.
\newblock \href
  {http://dx.doi.org/http://dx.doi.org/10.1016/j.energy.2009.11.023}
  {\path{doi:http://dx.doi.org/10.1016/j.energy.2009.11.023}}.
\newline\urlprefix\url{http://www.sciencedirect.com/science/article/pii/S036054420900512X}

\bibitem{14LWWSTHM}
H.~Lund, S.~Werner, R.~Wiltshire, S.~Svendsen, J.-E. Thorsen, B.~M.
  F.~Hvelplund, 4th generation district heating integration smart thermal grids
  into future sustainable energy systems, Energy 68 (2014) 1--11.

\bibitem{85MK}
A.~P. Merenkov, V.~Y. Khasilev, Theory of hydraulic circuits [in {R}ussian],
  Nauka, Moscow, 1985.

\bibitem{90SS}
V.~Sidler, Z.~Shalaginova, Mathematical models of the systems of heat delivery,
  in: Contemporary {P}roblems of {S}ystem {S}tudies in {P}ower {E}ngineering
  [in Russian], Irkutsk, 1990.

\bibitem{06GBLS}
I.~Gabrielaitine, B.~Bohm, H.~Larse, B.~Sunden, Dynamic performance of district
  heating system in {M}adum-vej, {D}enmark, 10th {I}nternational {C}onference
  on {D}istrict {H}eating and {C}ooling, 2006.

\bibitem{07GBS}
I.~Gabrielaitiene, B.~Bohm, B.~Sunden, Modelling temperature dynamics of a
  district heating system in {N}aestved, {D}enmark— a case study, Energy
  Conversion and Management 48~(1) (2007) 78 -- 86.

\bibitem{08GBS}
I.~Gabrielaitiene, B.~Bohm, B.~Sunden, Evaluation of {A}pproaches for
  {M}odeling {T}emperature {W}ave {P}ropagation in {D}istrict {H}eating
  {P}ipelines, {H}eat {T}ransfer {E}ngineering 29 (2008) 45–--56.

\bibitem{08NN}
N.~Novitsky, Mathematical models and solution methods of direct and inverse
  problems in dynamics of heat flux front in hydraulic networks, in: S.~e.
  Averyanov, Novistky (Ed.), Pipe Energy Systems. Development of theory and
  methods of mathemtical modeling and optimization. Seminar report from
  2006.[in Russian], Novosibirsk, Nauka, 2008.

\bibitem{09SZPMKT}
V.~D. Stevanovic, B.~Zivkovic, S.~Prica, B.~Maslovaric, V.~Karamarkovic,
  V.~Trkulja,
  \href{http://www.sciencedirect.com/science/article/pii/S0196890409001654}{Prediction
  of thermal transients in district heating systems}, Energy Conversion and
  Management 50~(9) (2009) 2167 -- 2173.
\newblock \href
  {http://dx.doi.org/http://dx.doi.org/10.1016/j.enconman.2009.04.034}
  {\path{doi:http://dx.doi.org/10.1016/j.enconman.2009.04.034}}.
\newline\urlprefix\url{http://www.sciencedirect.com/science/article/pii/S0196890409001654}

\bibitem{17SD}
K.~Sartor, P.~Dewalef,
  \href{http://www.sciencedirect.com/science/article/pii/S0360544217303444}{Experimental
  validation of heat transport modelling in district heating networks},
  Energy\href
  {http://dx.doi.org/http://dx.doi.org/10.1016/j.energy.2017.02.161}
  {\path{doi:http://dx.doi.org/10.1016/j.energy.2017.02.161}}.
\newline\urlprefix\url{http://www.sciencedirect.com/science/article/pii/S0360544217303444}

\bibitem{98Nov}
N.~Novitsky, Estimation of {P}arameters of {H}ydraulic {C}hains. [in Russian],
  Nauka, Novosibirsk, 1998.

\bibitem{16NV}
N.~Novitsky, O.~Vanteyeva, Probabilistic modeling of temperature conditions in
  the pipeline networks, in: A.~Karagrigoriou, T.~Oliveira, C.~Skiadas (Eds.),
  Statistical, {S}tochastic and {D}ata {A}nalysis {M}ethods and {A}pplications,
  ISAST; 1st edition, 2016, pp. 497--514.

\bibitem{13BLSZZ}
D.~Bertsimas, E.~Litvinov, X.~A. Sun, J.~Zhao, T.~Zheng, Adaptive robust
  optimization for the security constrained unit commitment problem, IEEE
  Transactions on Power Systems 28~(1) (2013) 52--63.

\bibitem{14BCH}
D.~Bienstock, M.~Chertkov, S.~Harnett,
  \href{http://dx.doi.org/10.1137/130910312}{Chance-{C}onstrained {O}ptimal
  {P}ower {F}low: {R}isk-{A}ware {N}etwork {C}ontrol under {U}ncertainty}, SIAM
  Review 56~(3) (2014) 461--495.
\newblock \href {http://arxiv.org/abs/http://dx.doi.org/10.1137/130910312}
  {\path{arXiv:http://dx.doi.org/10.1137/130910312}}, \href
  {http://dx.doi.org/10.1137/130910312} {\path{doi:10.1137/130910312}}.
\newline\urlprefix\url{http://dx.doi.org/10.1137/130910312}

\bibitem{15MN}
E.~Mikhailovsky, N.~Novitsky,
  \href{http://www.sciencedirect.com/science/article/pii/S2405722315300207}{A
  modified nodal pressure method for calculating flow distribution in hydraulic
  circuits for the case of unconventional closing relations}, St. Peterburg
  Polythechnical University Jouranl: Physics and Mathematics 1.
\newline\urlprefix\url{http://www.sciencedirect.com/science/article/pii/S2405722315300207}

\bibitem{95Fri}
U.~Frisch, Turbulence: {T}he legacy of {A}.{N}. {K}olmogorov, Cambridge
  University Press, 1995.

\bibitem{Shukhov}
Vladimir {S}hukhov, \url{https://en.wikipedia.org/wiki/Vladimir_Shukhov}.

\bibitem{85AKU}
B.~Agapein, B.~Krivoshein, B.~Ufin, Heat and hydraulic computations for oil and
  oil-product pipes [in {R}ussian], Nedra/Subsoil, Moscow, 1985.

\bibitem{82Kri}
B.~Krivoshein, Thermo-physical {C}omputations in {G}as {P}ipes [in {R}ussian],
  Nedra/Subsoil, Moscow, 1982.

\bibitem{75Sok}
E.~Y. Sokolov, District {H}eating and {D}istrict {H}eating {P}ipes [in
  {R}ussian], Energy, Moscow, 1975.

\bibitem{82Ion}
A.~Ionin, Heat {S}upply [in {R}ussian], Stroiizdat, Moscow, 1982.

\bibitem{86Zin}
N.~Zinger, {H}ydraulic and {H}eating {R}egimes of {D}istrict {H}eating
  {N}etworks [in {R}ussian], Energoatomizdat, Moscow, 1986.

\bibitem{88BVG}
I.~Belyakina, V.~Vitalev, N.~Gromov, Water {H}eating {S}ystems, in: N.~Gromov,
  E.~Shubina (Eds.), Handbook in {D}esign of {H}eating {S}ystems[in {R}ussian],
  Energoizdat, Moscow, 1988.

\bibitem{11CH}
D.~Callaway, I.~Hiskens, Achieving controllability of electric loads,
  Proceedings of the IEEE 99~(1) (2011) 184--199.
\newblock \href {http://dx.doi.org/10.1109/JPROC.2010.2081652}
  {\path{doi:10.1109/JPROC.2010.2081652}}.

\bibitem{14Siano}
P.~Siano,
  \href{http://www.sciencedirect.com/science/article/pii/S1364032113007211}{Demand
  response and smart grids—a survey}, Renewable and Sustainable Energy
  Reviews 30 (2014) 461 -- 478.
\newblock \href
  {http://dx.doi.org/http://dx.doi.org/10.1016/j.rser.2013.10.022}
  {\path{doi:http://dx.doi.org/10.1016/j.rser.2013.10.022}}.
\newline\urlprefix\url{http://www.sciencedirect.com/science/article/pii/S1364032113007211}

\bibitem{15OEB}
Ondeck, T.~Edgar, M.~Baldea, Optimal operation of a residential district-level
  combined photovoltaic/natural gas power and cooling system, Applied Energy
  156 (2015) 593--606.

\end{thebibliography}

\end{document}